\documentclass[12pt]{iopart}

\usepackage{graphicx}
\usepackage{amssymb}
\usepackage{amsfonts}
\usepackage{color}

\begin{document}

\title{Quantum-well and modified image-potential states in thin Pb(111) films: an estimate for the local work function}

\author{A Yu Aladyshkin$^{1,2}$}

\address{$^1$ Institute for Physics of Microstructures RAS, 603950, Nizhny Novgorod, GSP-105, Russia \\
$^2$ Lobachevsky State University of Nizhny Novgorod, 603950, Nizhny Novgorod, 23 Gagarin avenue, Russia}

\begin{abstract}
Quantum--confined electronic states such as quantum--well states (QWS) inside thin Pb(111) films and modified image--potential states (IPS) above the Pb(111) films grown on Si(111)7$\times$7 substrate were studied by means of low--temperature scanning tunnelling microscopy (STM) and spectroscopy (STS) in the regime of constant current $I$. By plotting the position of the $n-$th emission resonances $U^{\,}_n$ versus $n^{2/3}$ and extrapolating the linear fit for the dependence $U^{\,}_n(n^{2/3})$ in the high-$n$ limit towards $n=0$, we estimate the local work function for the Pb(111) film: $W\simeq 3.8\pm 0.1\,$eV. We experimentally demonstrate that modifications of the shape of the STM tip can change the number of the emission peaks associated with the resonant tunnelling via quantized IPS levels for the same Pb terrace; however it does not affect the estimate of the local work function for the flat Pb terraces. We observe that the maxima in the spectra of the differential tunnelling conductance $dI/dU$ related to both the QWS and the modified IPS resonances are less pronounced if the STM tip becomes more blunt.
\end{abstract}

\ead{aladyshkin@ipmras.ru}


\section{Introduction}

\hspace*{0.6cm} Ultra-thin metallic films and nanoislands appear to be one of the most suitable objects for the investigation of both transverse and lateral quantization effects in solid--state nanostructures.

\begin{figure*}[t!]
    \centering{\includegraphics[width=17cm]{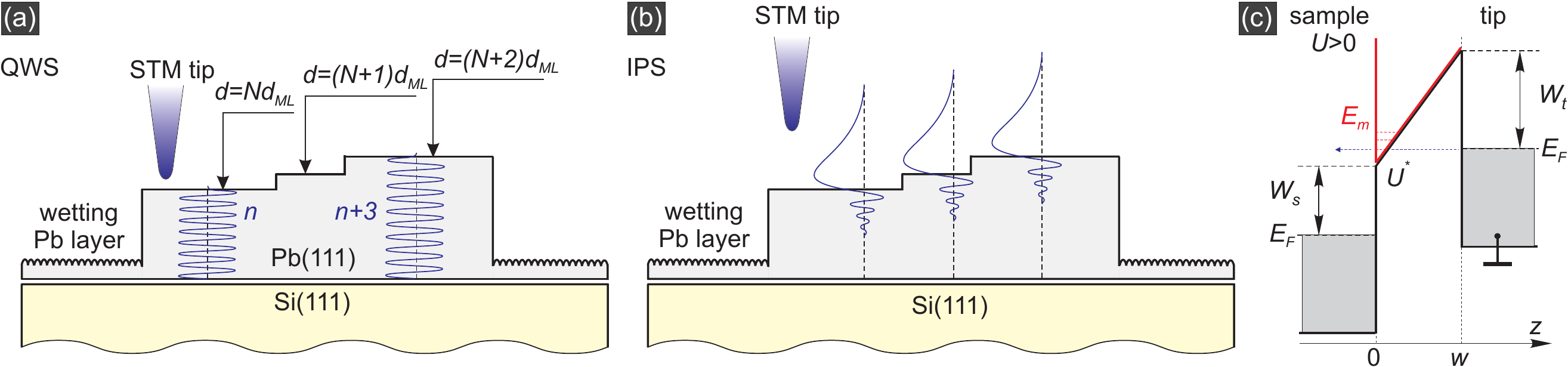}}
    \caption{(color online) {\bf (a)} Schematic representation of the QWS wave functions for Pb terraces of certain parity $N$ for an energy $E^*$ close to $E^{\,}_F$, where $N$ is the number of the Pb monolayers and the parameter $n$ characterizes the number of the electronic half--waves \cite{Putilov-JETPLett-19}. {\bf (b)} Schematic representation of the electronic wave functions for the ground state ($n=1$) of the modified IPS, which appear at each flat terrace. {\bf (c)}  Schematic energy diagram illustrating the field emission resonance via the modified IPS, here $w$ is the separation between the sample surface (collector) and the tip (emitter). Here $E^{\,}_F$ is the Fermi energy, $W^{\,}_s$ and $W^{\,}_t$ are the work functions for the sample and the tip, correspondingly.}
    \label{Fig01}
    \end{figure*}

Quantum--well states (QWS) are associated with the formation of standing waves for nearly-free electrons localized in a one-dimensional potential well formed by metal--substrate and metal--vacuum interfaces similar to a 'particle-in-a-box' problem \cite{Landau-III}. The wave functions corresponding to the QWS are shown schematically in Fig.~\ref{Fig01}a. The QWS in metallic films have been intensively studied by means of scanning tunnelling microscopy and spectroscopy (STM/STS) and by photoemission spectroscopy for the following systems: Pb/Si(111) and Pb/Si(111)$7\times 7$ \cite{Altfeder-PRL-97,Altfeder-PRL-02,Su-PRL-01,Eom-PRL-06,Hong-PRB-09,Dil-PRB-06,Mans-PRB-02, Ricci-PRB-09,Ustavshchikov-JETPLett-2017,Putilov-JETPLett-19}, In/Si(111) \cite{Dil-PRB-06,Altfeder-PRL-04}, Ag/Fe(100), Ag/V(100) and Cu/Co(100) \cite{Milun-RPP-02}, Ag/Au(111), Ag/Cu(111) and Ag/Ni(111) \cite{Chiang-SSR-00}.

Electronic states localized near the flat surfaces of crystals or surface states (SS) were predicted theoretically by Tamm \cite{Tamm-JETP-32} and Shockley \cite{Shockley-JETP-39}. They demonstrated that due to a discontinuity of the potential energy the stationary wave function could decay exponentially both in vacuum and in the crystal (with oscillations), provided that the energy of such localized state corresponds to a band gap of the bulk crystal. In reality, the electric potential outside the metallic crystal is represented by a slowly decaying function ($\sim 1/z$, where $z$ is the distance from the metal--vacuum interface) due to the electrostatic Coulomb interaction between the electron and its mirror charge. This makes it possible to trap electrons in the potential well formed by the surface of the crystal from one side and the Coulomb potential from the other side (see reviews \cite{McRae-RMP-79,Echenique-ProgSurfSci-90} and references therein). Such electronic states are known as image potential states (IPS) and they correspond to localization of electrons {\it above} the conducting surface. Usually the IPS in the absence of external electric fields are studied by means of photoemission spectroscopy \cite{Garcia-PRL-85,Hofer-Science-97}.

The effect of electron localization in a nonuniform electric field becomes more pronounced in the presence of external sources, what is typically encountered in the STM technique. Indeed, for a non-zero potential difference (bias) $U$ between the STM tip and the sample, the electric potential near the surface appears to be a superposition of the potential of the image charge and the linearly increasing function of $z$. Obviously, the positions of all IPS in the presence of the external electric field are shifted upward to higher energies similar to the Stark effect for a hydrogen-like atom \cite{Landau-III}. We will use the term 'modified IPS' to refer to the Stark--shifted IPS, whose energy spectrum is mainly controlled by the linearly increasing electric potential. The wave functions corresponding to the ground state of the modified IPS are shown schematically in Fig.~\ref{Fig01}b. The influence of the quantized electronic states localized in the triangular potential well on the energy dependence of the transmission coefficient for the trapezoidal tunnelling barrier and on the current--voltage ($I-U$) dependence were theoretically considered by Gundlach \cite{Gundlach-SolStateElectr-66}. Sometimes the resulting oscillations of the differential conductance $dI/dU$ as a function of $U$ are referred to as the Gundlach oscillations. The field emission resonances attributed with the resonant tunnelling through the quasistationary modified IPS were experimentally studied for the following atomically flat surfaces: Au(110) \cite{Becker-PRL-85}, Ni(100) \cite{Binnig-PRL-85}, surface of diamond C(100)$2\times 1$ \cite{Bobrov-Nature-01}, Cu(100) \cite{Wahl-PRL-03}, Mo(110) \cite{Jung-PRL-95}, Ag(100) and Fe(110) \cite{Hanuschkin-PRB-07}, Fe(110) \cite{Kubetzka-APL-07}, grahene \cite{Silkin-PRB-09}, InAs(111) \cite{Martinez-Blanco-PRB-15}, topological semimetal Sb(111) \cite{Ge-PRB-20}; and for nanostructured objects on top of flat surfaces:  FeO islands on Pt(111) \cite{Rienks-PRB-05}, molecules of benzene on Cu(111) \cite{Dougherty-PRL-06}, NaCl islands on Ag(100) \cite{Pivetta-PRB-05,Ploigt-PRB-07}, Ag islands on Au(111), Cu(111) and Co islands on Cu(111)\cite{Lin-PRL-07}, Na islands on Cu(111) \cite{Borisov-PRB-07}, Pb islands on Cu(111) and Ag(111) \cite{Yang-PRL-09,Becker-PRB-10} with theoretical interpretation \cite{Zugarramurdi-PRB-11}, Co islands on Au(111) \cite{Schouteden-PRL-09}, surface defects like stacking-fault tetrahedrons on Ag(111) \cite{Schouteden-PRL-12}, carbon nanotubes on Au(111) \cite{Schouteden-Nanotech-10}), graphene islands on Ir(111) \cite{Craes-PRL-13}), and others. In particular, the analysis of the spectra of the IPS resonances allows to estimation the local work function, {\it e.g.} for NaCl \cite{Pivetta-PRB-05,Ploigt-PRB-07}, Ag \cite{Lin-PRL-07}, Pt \cite{Kolesnychenko-RSI-99,Kolesnychenko-PhysB-00}. To the best of our knowledge, there are no reports concerning the higher-order Gundlach oscillations in thin Pb(111) films on top of reconstructed Si(111)$7\times 7$ surface and the estimates of the local work function for such system.

In this paper we present the results of the experimental investigations of the QWS-- and the modified IPS resonances in thin Pb(111) films with flat terraces on Si(111)$7\times 7$ surface by means of low--temperature STM/STS in the regime of constant tunnelling current $I$ and variable distance $z$ between the STM tip and the surface. Maxima of the differential tunnelling conductance $dI/dU$ and the rate of height variations $dz/dU$ at $U\lesssim 3\,$V can be associated with the QWS inside the Pb film and their energies are sensitive to the local thickness of the Pb film. Indeed, the change in the Pb thickness by one monolayer converts the local $dI/dU$ and $dz/dU$ maxima into local minima and vice versa. At higher voltages ($U\gtrsim 3.5\,$V), the maxima of $dI/dU$ and $dz/dU$ become independent from the Pb thickness and, therefore, such the resonances should be attributed to the modified IPS. We demonstrate that the abrupt change in the shape of the STM tip results in an increase in the number of the IPS resonances for the same Pb terrace. We show that neither the change in the local Pb thickness nor the modification of the STM tip affects the estimate of the local work function $W\simeq 3.8\pm 0.1\,$eV. We observe that the changes of the tip shape could make the QWS resonances less pronounced without affecting the resonant energies. These findings reveal the importance of the particular shape of the tip for the visible form of the tunnelling spectra \cite{Comment}.

\section{Experimental procedure}

\hspace*{0.6cm} Investigation of the structural and electronic properties of Pb nanostructures was carried out on an ultra-high vacuum (UHV) low-temperature scanning probe microscopy (SPM) setup by Omicron Nanotechnology GmbH. Thermal deposition of Pb was performed on the reconstructed Si(111)7$\times$7 surface at room temperature {\it in-situ} in UHV conditions (see other details in Refs. \cite{Ustavshchikov-JETPLett-2017,Putilov-JETPLett-19}).

The topography of the Pb islands was then studied by STM at 78~K in the regime of constant tunnelling current with a modulated bias potential of the sample $U=U^{\,}_0+U^{\,}_1\,\cos(2\pi f^{\,}_0t)$, where $f^{\,}_0=7285$~Hz and $U^{\,}_1=40\,$mV. Provided that $f^{\,}_0$ significantly exceeds the threshold frequency of the feedback loop ($\sim200$~Hz), the modulated potential of the sample should not result in the appearance of any artifacts on the topographic images. In order to modify the shape of the W tips {\it in-situ}, we routinely touch the surface of clear Pb terraces by changing the distance $z$ between the tip and the sample surface from zero (current position) to $-3\,$nm (well below surface) during a standard procedure of local $I-z$ spectroscopy. As a result, the W tips were covered by a Pb layer.

    \begin{figure*}[b!]
    \centering{\includegraphics[width=16cm]{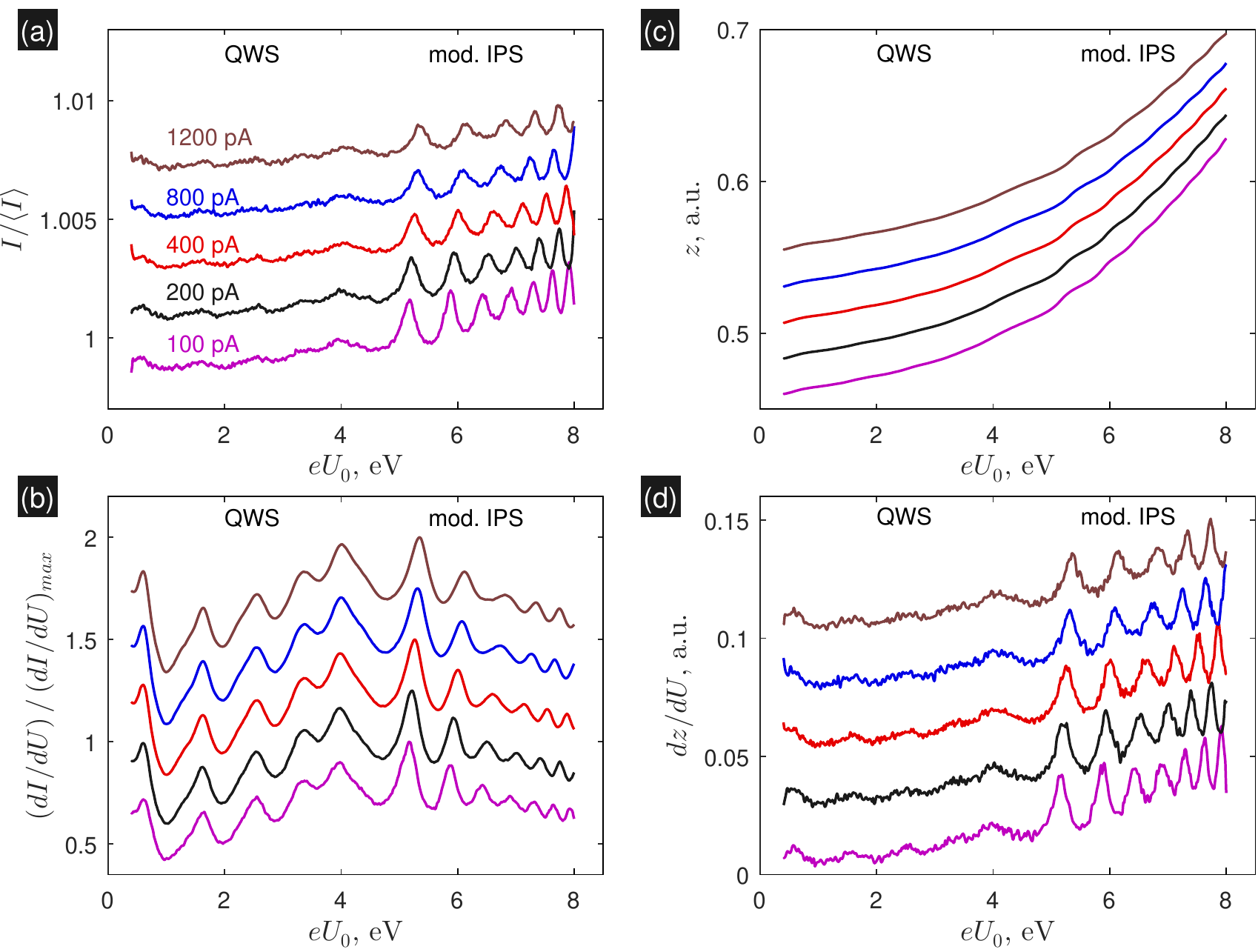}}
    \caption{(color online) {\bf (a--d)} Typical dependences of the normalized tunnelling current $I$ (a), the normalized differential conductance $dI/dU$ (b), the nominal height of the tip $z$ (c), and the rate of the height variation $dz/dU$ on the mean bias $U^{\,}_0$ for the case of thin Pb island on the Si(111)$7\times 7$ surface at 78 K and the same location (the local Pb thickness is about 3.7\,nm). All measurements were carried out in the regime of the constant mean tunnelling current for $\langle I \rangle = 100, 200, 400, 800$ and 1200 pA (from bottom to top), all the curves are shifted vertically for clarity.}
    \label{Fig07}
    \end{figure*}

    \begin{figure*}[t!]
    \centering{\includegraphics[width=17cm]{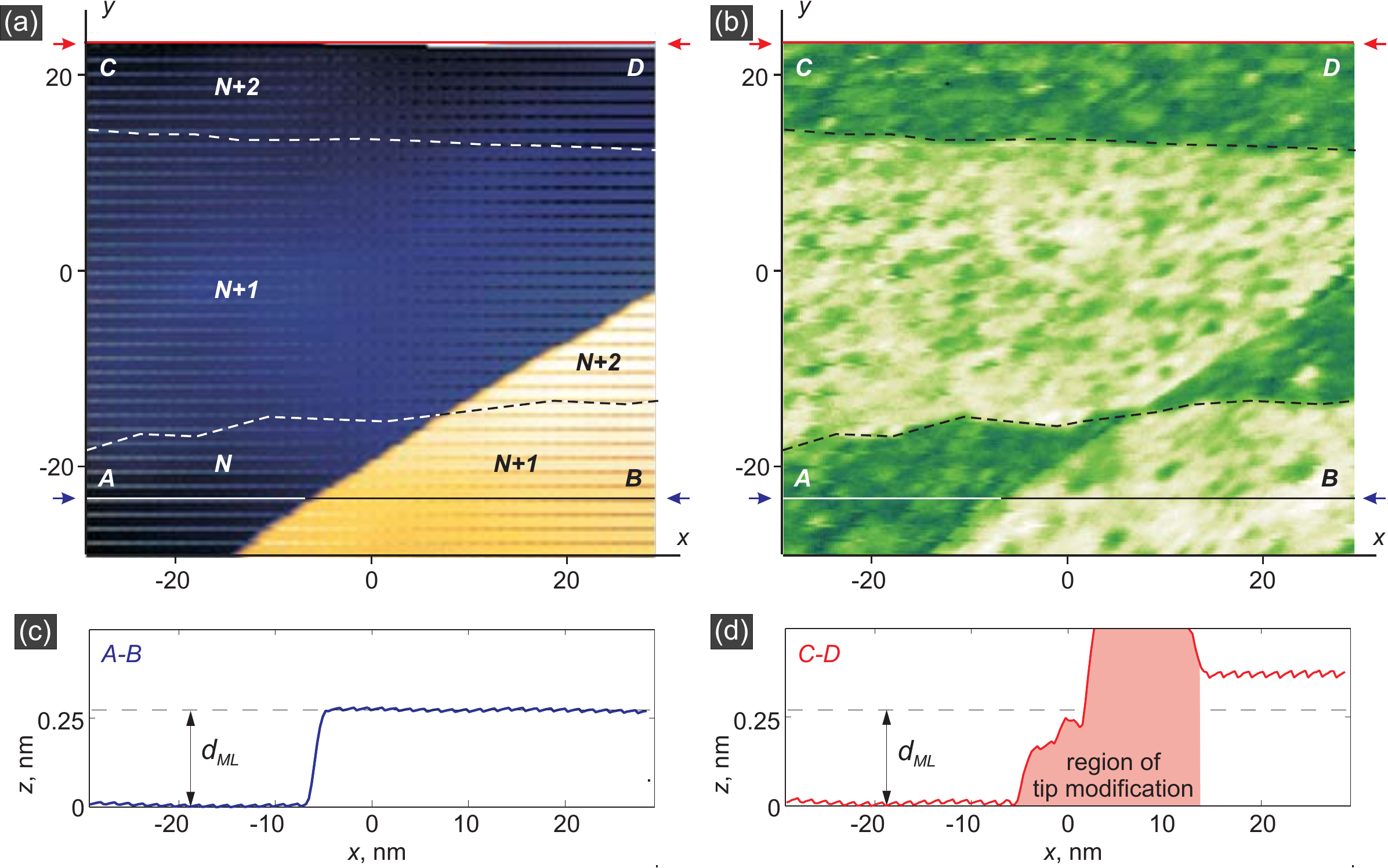}}
    \caption{(color online) {\bf (a, b)} The topography map (a) and the map of the differential tunnelling conductance $dI/dU$ (b) for the Pb island on Si(111)$7\times 7$, acquired simultaneously during scanning STS measurements over a grid of 40$\times$37 points in the regime of constant current (image size is 58$\times$54\,nm$^2$, $U^{\,}_0=+500$\,mV, $\langle I\rangle=800$\,pA, $T=78\,$K). The estimated thickness $d$ of the island in the lower left corner is 13.4\,nm or $N=d/d^{\,}_{ML}\simeq 47$; the dashed lines indicate the positions of monoatomic step in the Si substrate. Lighter shades on the $dI/dU$ map correspond to higher conductance values and vice versa. {\bf (c)} Profile $z(x)$ along the $A-B$ line, the variation in height is equal to $d^{\,}_{ML}$, the visible corrugation is a consequence of the grid spectroscopical measurements. {\bf (d)} Profile $z(x)$ along the $C-D$ line, the variation in height for the same Pb terrace is caused by the abrupt modification of the tip.}
    \label{Fig02}
    \end{figure*}

The electronic properties of the Pb islands were subsequently investigated by STS in the regime of constant tunnelling current and variable $z$. In particular, we study the dependence of the amplitude of the first Fourier component of the current $I^{\,}_1$ on $U^{\,}_0$ as a function of location $(x,y)$ in the lateral plane. Under the condition $U^{\,}_1\ll U^{\,}_0$, the $I^{\,}_1$ value measured by a Stanford Research SR\,830 lock--in amplifier is proportional to the differential tunnelling conductance $dI/dU$ at $U=U^{\,}_0$. Simultaneously, we obtain the dependence of the tip--sample distance $z$ on $U^{\,}_0$ by recording the signal of the feedback loop. We regularize the measured dependences $z(U^{\,}_0)$ in order to remove high--frequency noise using a running Gaussian filter with a window of 40 mV and then numerically calculate the rate of the height variation $dz/dU$ as a function of $U^{\,}_0$. We would like to emphasize that the modulation scanning tunnelling technique allows us to acquire the topographic map and a 2D map of the differential conductance $dI/dU(x,y)$ simultaneously. The $dI/dU-$maps facilitate the determination the constant height areas and the location of the monatomic steps on the substrate and the hidden defects \cite{Ustavshchikov-JETPLett-2017}.

\section{Results and discussion}

\hspace*{0.6cm} Figure~\ref{Fig07} shows typical local spectra for thin Pb nanoisland measured at the same location in the regime of constant tunnelling current and variable tip--sample distance $z$. The local Pb thickness is about 3.7\,nm or $13\,d^{\,}_{ML}$, where $d^{\,}_{ML}=0.285\,$nm is the thickness of a Pb monolayer for the Pb(111) surface. It is easy to see that there are two clearly distinctive types of oscillations at low and high $U^{\,}_0$ values. The peaks of the dependence of the differential tunnelling conductance $dI/dU$ at low $U^{\,}_0$ shown in Fig.~\ref{Fig07}b should be attributed to the resonant tunnelling through the QWS, what is consistent with data in Refs. \cite{Altfeder-PRL-97,Altfeder-PRL-02,Su-PRL-01,Eom-PRL-06,Hong-PRB-09,Ustavshchikov-JETPLett-2017,Putilov-JETPLett-19}. It is interesting to note that tunnelling through the QWS is accompanied by the appearance of poorly defined maxima in the dependences of the tunnelling current $I$ and the rate of the height variations $dz/dU$ on $U^{\,}_0$. For this reason we plot $\log dz/dU$ instead of $dz/dU$ and analyze the tunneling conductance spectra ({\it i.e.} the dependences $dI/dU$ on $U^{\,}_0$) to extract the resonant QWS values. The position of the QWS resonances are practically independent on the measuring current up to 1200 pA. In contrast to that, the oscillatory behavior related to the tunnelling through the modified IPS is clearly seen in all presented dependences $I$, $dI/dU$ and $dz/dU$ on $U^{\,}_0$ in the high $U^{\,}_0$ limit (panels a, b and d in Fig.~\ref{Fig07}). It should be emphasized that the positions of the IPS resonances depend on the initial height and thus on the measuring current $\langle I \rangle$, what can be considered as a simple practical way to distinguish the QWS from the IPS resonances. However due to relatively small thickness of the studied Pb film, the typical periods and amplitudes of the QWS- and IPS-oscillations are rather close, therefore it is hard to define the threshold value of $U^{\,}_0$ separating two types of oscillations for the data shown in Fig.~\ref{Fig07}.

Figure~\ref{Fig02}a shows the topography $z(x,y)$ of a thicker Pb island with two flat terraces on the upper surface. The map of the differential conductance $dI/dU(x,y)$ shown in Fig.~\ref{Fig02}b is given as evidence of the presence of two monatomic steps on the Si substrate \cite{Ustavshchikov-JETPLett-2017}. The height profile $z(x)$ along the $A-B$ line (Fig.~\ref{Fig02}c) also indicates that the variation of the local height is equal to $d^{\,}_{ML}$. The interpretation of the pseudo-colored $dI/dU$ map is rather simple. Due to the peculiar relationship $\lambda^{\,}_F/d^{\,}_{ML}\simeq 4/3$ \cite{Su-PRL-01} between the Fermi wavelength $\lambda^{\,}_F\simeq 0.394\,$nm and $d^{\,}_{ML}$ \cite{Su-PRL-01,Eom-PRL-06,Hong-PRB-09,Ustavshchikov-JETPLett-2017}, one can conclude that three electronic half--waves at the Fermi level is close to the thickness of two monolayers. It means that if there is a QWS, characterized by the eigenenergy $E^*$ (close to the Fermi energy $E^{\,}_F$) and by the number of the half-waves $n$, for the terrace of the thickness of $N$ monolayers, then the QWS with the same $E^*$ and $n+3$ should appear for the terrace of the thickness of $N+2$ monolayers. Consequently, the local differential conductance should have the local maxima at $eU\simeq E^*-E^{\,}_F$ for the terraces of the same parity ($N$ and $N\pm 2$) and the local minima for the terraces of opposite parity ($N\pm 1$).

We would like to emphasize that at the end of such time-consuming experiment (about 10 hours) some parts of the tip are worn due to unavoidable repeated contact with the sample surface. This causes eventually unpredictable changes in the shape of the tip. This conclusion is confirmed by the last line profile along the $C-D$ line via the area of this local catastrophe as evidenced by the anomalous thickness variation (Fig.~\ref{Fig02}d). Interestingly, this unwanted tip damage/modification allows us to compare the local electronic properties of the Pb island for two important cases: (i) two terraces of different height ($N$ and $N+1$ monolayers) tested by the same STM tip, and (ii) two areas on the same terrace ($N+2$ monolayers) tested by the STM tip before and after the modification.

    \begin{figure*}[ht]
    \centering{\includegraphics[width=16cm]{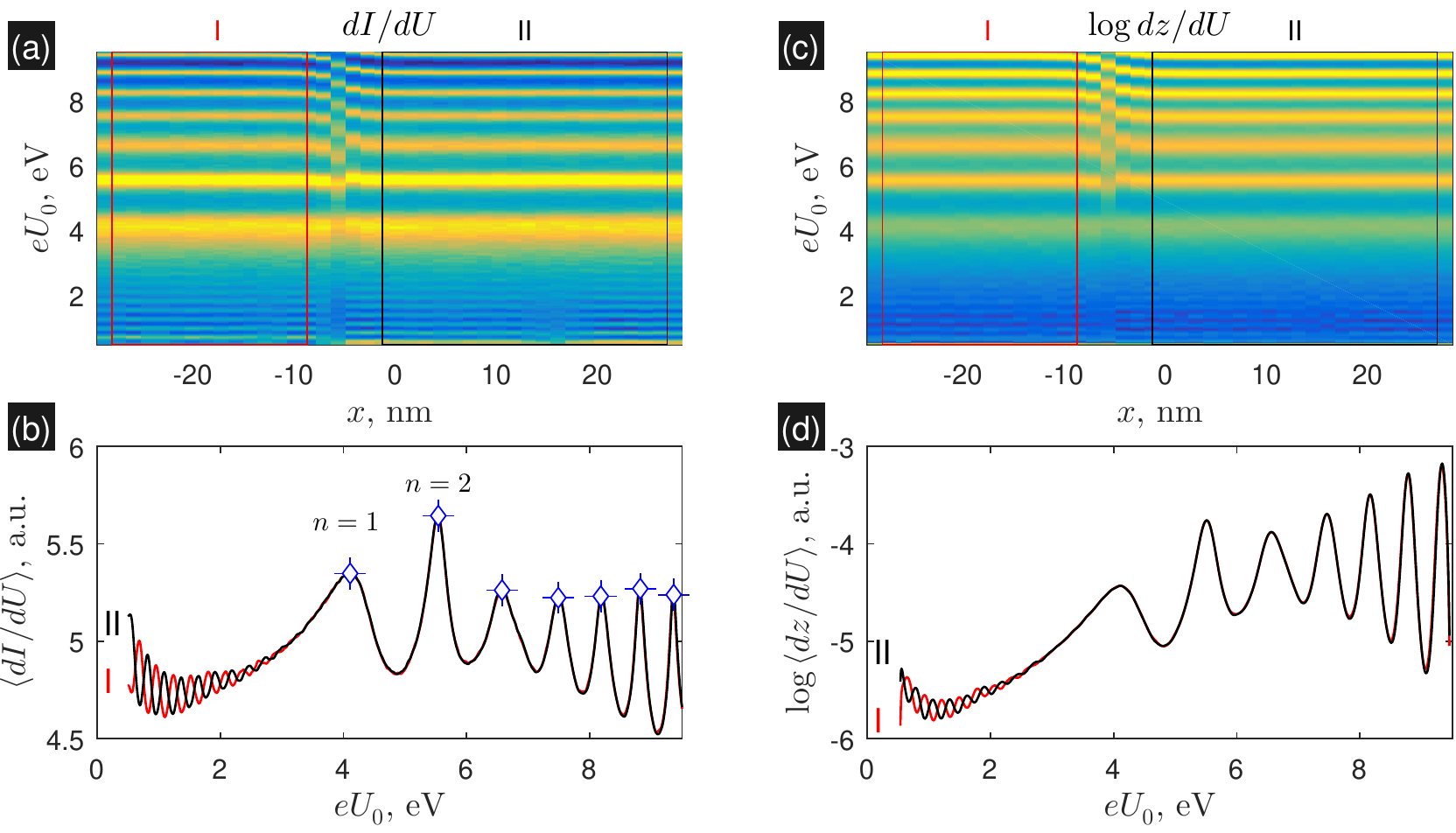}}
    \caption{(color online) {\bf (a, c)} Dependence of the differential tunnelling conductance $dI/dU$ (a) and the rate of height variation $\log\,dz/dU$ (c) on the bias $U^{\,}_0$ and the $x-$coordinate along the $A-B$ line (see Fig.~\ref{Fig02}), measured simultaneously at given current $\langle I\rangle = 800\,$pA. {\bf (b, d)} Dependences of $\langle dI/dU \rangle$ and $\log\,\langle dz/dU \rangle$ on $U^{\,}_0$, averaged over $x-$coordinate within areas I and II.}
    \label{Fig03}
    \end{figure*}

    \begin{figure*}[th!]
    \centering{\includegraphics[width=16cm]{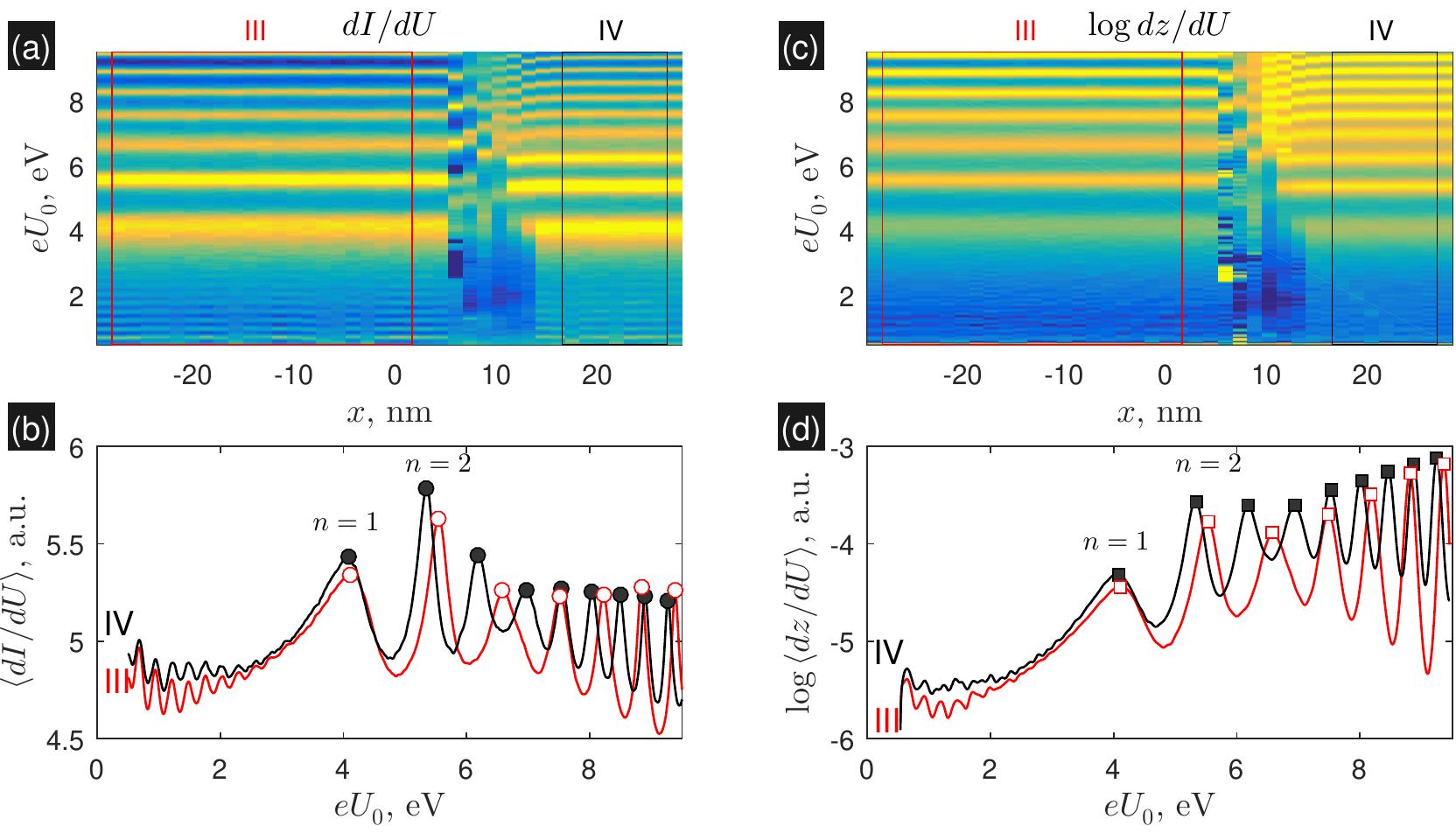}}
    \caption{(color online)
    {\bf (a, c)} Dependence of the differential tunnelling conductance $dI/dU$ (a) and the rate of height variation $\log\,dz/dU$ (c) on the bias $U^{\,}_0$ and the $x-$coordinate along the $C-D$ line (see Fig.~\ref{Fig02}), measured simultaneously at given current $\langle I\rangle = 800\,$pA. {\bf (b, d)} Dependences of $\langle dI/dU \rangle$ and $\log\,\langle dz/dU \rangle$ on $U^{\,}_0$, averaged over $x-$coordinate within areas III and IV.}
    \label{Fig04}
    \end{figure*}


Figure~\ref{Fig03} shows the color-coded maps illustrating the dependences of $dI/dU$ and $\log\,dz/dU$ on the mean bias potential $U^{\,}_0$ and the $x-$coordinate along the $A-B$ line (panels a and c, respectively). In order to facilitate further analysis, we plot the dependences $\langle dI/dU\rangle^{\,}_I$ and $\log\,\langle dz/dU\rangle^{\,}_I$ averaged over the $x-$coordinate within the region I as well as $\langle dI/dU\rangle^{\,}_{II}$ and $\log\,\langle dz/dU\rangle^{\,}_{II}$ averaged over the $x-$coordinate within the region II (panels b and d). We would like to point out that there are two types of oscillations with different specific periods and amplitudes. Almost equidistant small--scale oscillations on $dI/dU$ and $dz/dU$ at $U^{\,}_0\lesssim 3\,$V can be attributed to the resonant tunnelling through the QWS. Note that the change in the thickness by one monolayer drastically modifies the oscillatory dependence, as expected for the QWS in the Pb films. Taking the interval between two adjacent peaks $\Delta E\simeq 0.28\,$eV and the value of the Fermi velocity $v^{\,}_F\simeq 1.8\cdot 10^8\,$cm/sec \cite{Ustavshchikov-JETPLett-2017}, we can estimate the local thickness: $d\simeq \pi\hbar v^{\,}_F/\Delta E \simeq 13.4$\,nm or $N=d/d^{\,}_{ML}\simeq 47$. For $U^{\,}_0\gtrsim 3.5\,$V a different regime characterized by aperiodic large--scale oscillations on $dI/dU$ and $dz/dU$ emerges, which seem to be independent of the local thickness. This regime corresponds to the field emission resonances due to the resonant tunnelling through the modified IPS. In addition, we check that the positions of all IPS resonances expectedly shift to higher energies as the measurement current increases and, correspondingly, the initial height of the tip decreases (similar to Fig.~\ref{Fig07}).

Figure~\ref{Fig04} shows the spectroscopical results obtained along the $C-D$ line before and after the tip has been modified. The regions III and IV correspond to the same Pb terrace with the local thickness $N+2$, what can be verified by examining the position of the QWS resonances on the dependences of $\langle dI/dU\rangle^{\,}_{I}$, $\langle dI/dU\rangle^{\,}_{III}$, and $\langle dI/dU\rangle^{\,}_{IV}$ on $U^{\,}_0$. Since the number of the IPS resonances increases by two in the same energy interval, we conclude that the effective electrical field $F^*$ near the surface becomes weaker and, consequently, the tip radius increases. Sometimes we observe the decrease in the number of the IPS resonances indicating that the tip after a touching of the Pb surface and/or voltage pulses becomes sharper. Thus, the visualization of the IPS resonances can be a powerful tool to monitor the properties of the tip during STM/STS measurements. It should be mentioned that the amplitudes of both the low-energy QWS resonances and the high-energy IPS resonances on the $U^{\,}_0$ dependences of $\langle dI/dU\rangle^{\,}_{IV}$ and $\langle dz/dU\rangle^{\,}_{IV}$ become smaller after the tip modification. This points to the importance of the tip shape ({\it i.e.} the form of the potential barrier between the sample and the tip) for resolving the formation of the quantized subbands in thin Pb films on the basis of tunnelling spectroscopy measurements \cite{Comment}. We believe that our observations can be considered as a supporting argument for the concept of directional tunnelling \cite{Iavarone-PhysC-03}, in addition to pronounced anisotropy of the effective masses \cite{Dil-PRB-06,Altfeder-PRL-98} or localization effects \cite{Altfeder-PRL-04} proposed earlier for explaining the particular energy dependence of the tunnelling spectra in thin Pb films.

    \begin{figure}[t!]
    \centering{\includegraphics[width=8.5cm]{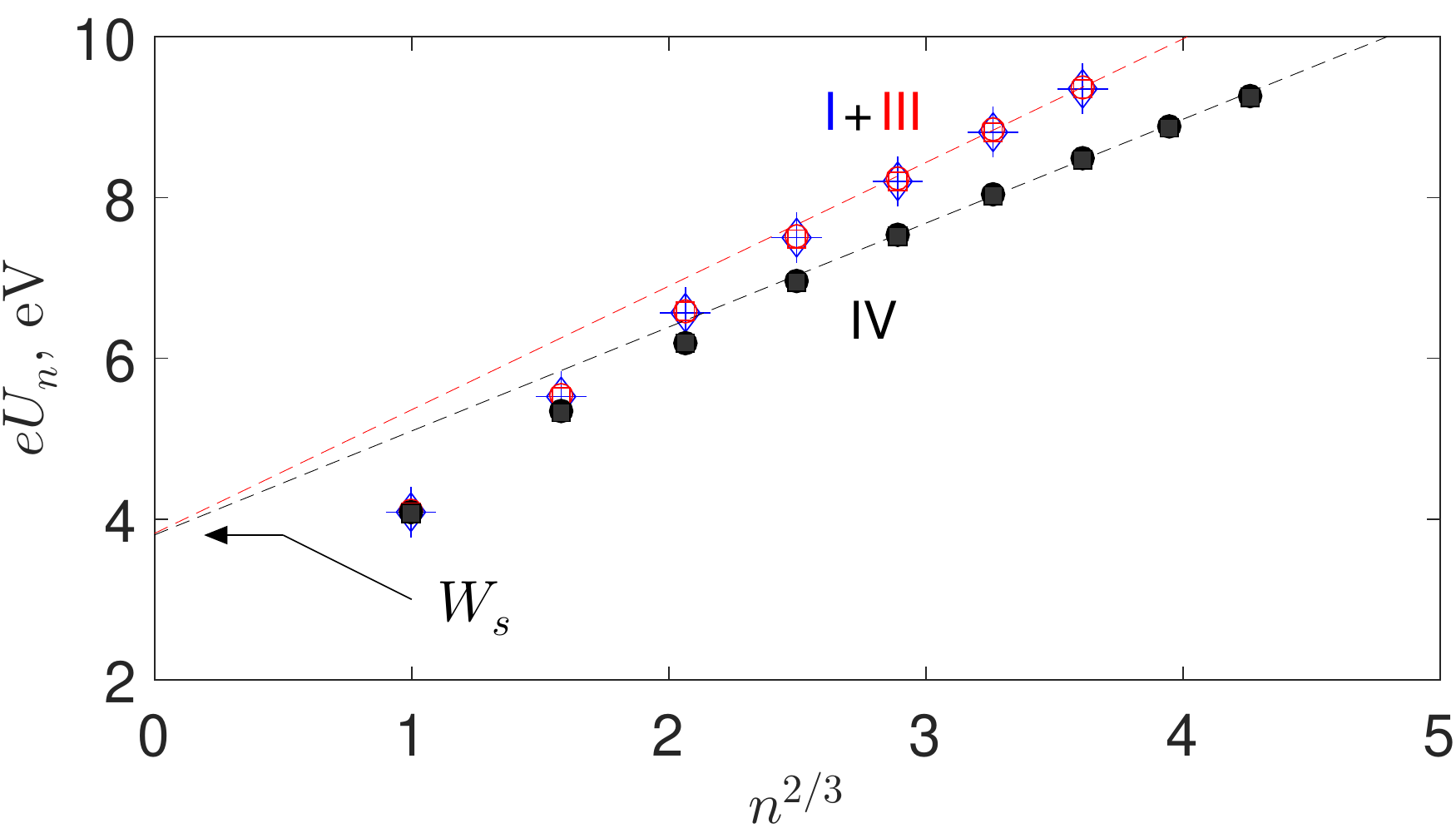}}
    \caption{(color online) Dependence of the energy $eU^{\,}_n=E^{\,}_n-E^{\,}_F$ of the $n-$th emission resonance on $n^{2/3}$: open symbols correspond to the maxima on the $\langle dI/dU\rangle^{\,}_{I}$ and $\langle dI/dU\rangle^{\,}_{III}$ vs. $U^{\,}_0$ dependences ($\textcolor[rgb]{0.00,0.07,1.00}{+}$ --- Fig.~\ref{Fig03}b, $\textcolor[rgb]{1.00,0.00,0.00}{\circ}$ --- Fig.~\ref{Fig04}b and $\textcolor[rgb]{1.00,0.00,0.00}{\square}$ --- Fig.~\ref{Fig04}d), filled symbols correspond to the maxima on the $\langle dI/dU\rangle^{\,}_{IV}$ vs. $U^{\,}_0$ dependences ($\bullet$ --- Fig.~\ref{Fig04}b and $\blacksquare$ --- Fig.~\ref{Fig04}d); the scattering interval for the $U^{\,}_n$ values is smaller than the symbol size. Extrapolation of the linear approximations for the  $U^{\,}_n(n^{2/3})$ dependences in the high $n-$limit towards $n=0$ gives us an estimate of the local work function: $W^{\,}_s\simeq 3.8\pm 0.1\,$eV.}
    \label{Fig05}
    \end{figure}

It is instructive to show how the relationship for the positions of the modified IPS resonances for two electrodes with flat surfaces can be reproduced from the Bohr--Sommerfeld quantization rule \cite{Landau-III}. We can neglect the contribution of the image potential and assume that the potential energy near the flat metallic surface is $U^{\,}_{pot}(z)=U^*+F^*z$ for $z>0$ and infinite for $z<0$ (see solid red line in Fig.~\ref{Fig01}c). Here $U$ is the electric potential of the sample with respect to the tip, $U^*=E^{\,}_F+W-|e|U$ is the bias--shifted energy of the bottom of the triangular potential well, $W^{\,}_s$ is the work function of the sample, $F^*=U/w + \Delta W$ is the effective electric field, including the effect of the Volta contact potential due to the difference $\Delta W=W^{\,}_t-W^{\,}_s$ in the work functions, and $w$ is the distance between the STM tip and the sample surface. We start with the substitution of $U^{\,}_{pot}(z)$ into the quantization rule
    \begin{eqnarray}
    \label{Bohr-Sommerfeld}
    \frac{1}{\pi\hbar}\,\int\limits_{a}^{b} \sqrt{2m^{\,}_0 (E^{\,}_{m}- U^{\,}_{pot}(z))}\,dz = m + \gamma,
    \end{eqnarray}
where $a=0$ and $b=(E-U^*)/F^{*}$ are classical turning points, $m^{\,}_0$ is the mass of free electron, $m=0, 1, \ldots$ is the integer index, the numerical coefficient $\gamma$ is close to 3/4 due to the presence of the impenetrable barrier at $z=0$ for the considered profile of the potential energy. After integration, we obtain the discrete energy spectrum for the electrons localized in the triangular potential well
    \begin{eqnarray}
    \label{Eq:IPS-triangular-well-2}
    \nonumber
    E^{\,}_{m} = E^{\,}_F + W^{\,}_s - |e|U + \left\{\frac{3}{2}\,\frac{\pi\hbar}{\sqrt{2m^{\,}_{0}}}\,F^{*}\,\left(m + \frac{3}{4}\right)\right\}^{2/3}.
    \end{eqnarray}
Since the process of resonant tunnelling starts when the quantized energy $E^{\,}_{m}$ in the area between the emitter (tip) and collector (sample) reaches the Fermi energy of the tip \cite{Ferry-book-09}, we come to the relationship
    \begin{eqnarray}
    \label{Eq:IPS-triangular-well-3}
    |e|U^{\,}_n \simeq W^{\,}_s + \left(\frac{3}{2}\,\frac{\pi\hbar}{\sqrt{2m^{\,}_{0}}}\,F^{*}\right)^{2/3}\,\left(n-\frac{1}{4}\right)^{2/3},
    \end{eqnarray}
where $n=1, 2, \ldots$ and the electric field $F^{*}$ is assumed to be constant. It is important to note, that this expression (\ref{Eq:IPS-triangular-well-3}) is identical to that derived for the positions of the $dI/dU$ maxima for the tunnelling junction with trapezoidal potential barrier between two metals in the free-electron-gas approximation \cite{Gundlach-SolStateElectr-66,Kolesnychenko-PhysB-00}. Surely, the positions of the modified IPS resonances should be lower than those predicted by Eq. (\ref{Eq:IPS-triangular-well-3}), due to the contribution of the image potential. Nevertheless, this effect should be small for the higher-order modified IPS resonances. Since $(n-1/4)^{2/3} \simeq n^{2/3}$ for $n\gg 1$, we conclude that (i) the energy of the $n-$th field emission resonance depends on the effective electrical field $F^{*}$ near the surface and on its number as $n^{2/3}$, leading to the decrease in the interval between two modified IPS resonances as $n$ increases; (ii) the extrapolation of the linear fitting function for the dependence $U^{\,}_n$ on $n^{2/3}$ at $n\gg 1$ towards $n=0$ gives us the estimate for the work function $W^{\,}_s$ regardless of the work function of the tip $W^{\,}_t$ and the actual separation $w$ between the tip and the sample surface. This approach was already used for estimating the work functions for Pt wire \cite{Kolesnychenko-PhysB-00}, and for Ag and Co nanoislands \cite{Lin-PRL-07}. More sophisticated methods for the extraction of the local work function for NaCl islands based on the comparison of the measured $dI/dU$ spectra with the results of numerical calculations were developed in Refs.~\cite{Pivetta-PRB-05,Ploigt-PRB-07}.

Figure~\ref{Fig05} shows the dependence of the position of the modified IPS resonances $U^{\,}_n$ on $n^{2/3}$ for the results presented in Fig.~\ref{Fig03}b and Fig.~\ref{Fig04}b,d. Both linear asymptotes of the $U^{\,}_n(n^{2/3})$ dependences for the high-$n$ limits intersect the axis of ordinates at the same point. This allows us to conclude that (i) the local work function for Pb film is close to $3.8 \pm 0.1\,$eV, and (ii) this estimate of the local work function does not depend on the local Pb thickness as well as the tip shape. Our estimate is in agreement with the theoretical value (3.85~eV) calculated for the Pb(111) surface \cite{Lang-PRB-71}.

    \begin{figure}[t!]
    \centering{\includegraphics[width=8.5cm]{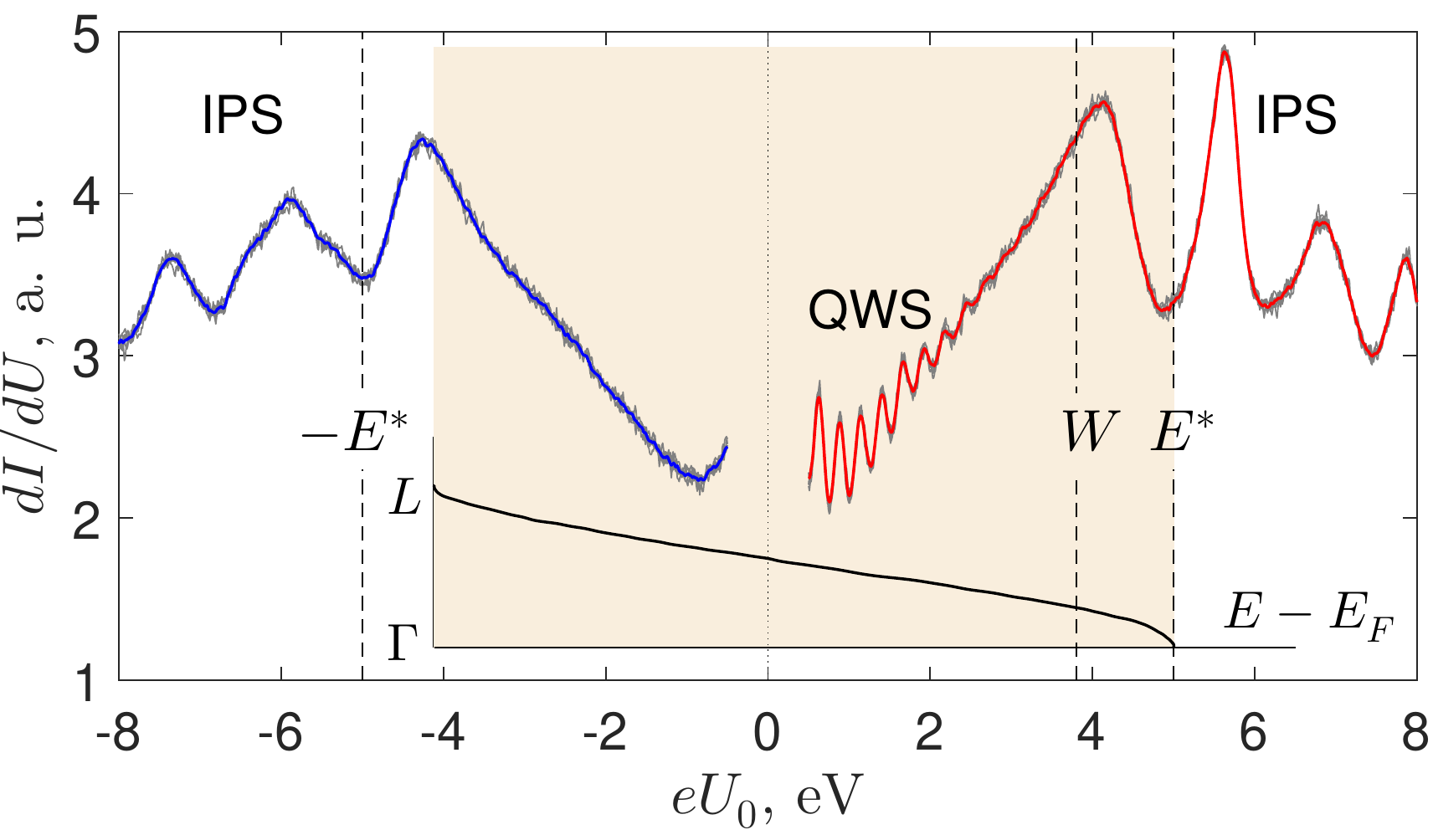}}
    \caption{(color online) Dependence of the differential tunnelling conductance $dI/dU$ on the mean bias $U^{\,}_0$, measured at $\langle I\rangle=-400\,$pA (left part, $U^{\,}_0<0$) and $\langle I\rangle=+400\,$pA (right part, $U^{\,}_0>0$) and averaged over five measurements for the same location. The electronic band structure $E(k^{\,}_z)$ along $\Gamma-L$ for $p^{\,}_z-$orbitals of bulk Pb crystal is taken from Refs.~\cite{Hong-PRB-09,Zugarramurdi-PRB-11}. Vertical dashed lines correspond to our estimate for the work function ($W^{\,}_s\simeq 3.8\,$eV) and to the top of the conduction band $\pm E^*$ ($E^*\simeq 5.0\,$eV).}
    \label{Fig06}
    \end{figure}

Finally, let us investigate the effect of the bias polarity on the shape of the dependence $dI/dU$ (Fig.~\ref{Fig06}). The pronounced asymmetry of the QWS resonances was explained by the difference in the transmission coefficient of the trapezoidal potential barrier for the positive and negative biases \cite{Altfeder-PRL-97}. The pronounced asymmetry of the tunnelling conductance spectrum at high $U^{\,}_0$ values reflects a different spatial structure of the electric field near the flat sample surface (at $U^{\,}_0>0$) or near the sharp STM tip endpoint (at $U^{\,}_0<0$), leading to the different number of the visible modified IPS resonances in the same range of the bias voltage. We compare the measured dependence $dI/dU$ on $U^{\,}_0$ with the electronic band structure $E(k^{\,}_z)$ calculated for bulk Pb crystal along $\Gamma-L$ direction \cite{Hong-PRB-09,Zugarramurdi-PRB-11}. Interestingly, the position of the $dI/dU$ minimum between the first and second IPS resonances for both polarities coincides with the energy of the top of the conduction band $E^*\simeq 5.0\,$eV. This explains the observed difference between the widths of the first broad IPS peak and other IPS resonances. Indeed, the IPS states with $n\ge 2$ are in the energy gaps for bulk Pb crystal, what suppresses the electron scattering and correspondingly, increases the lifetime of the higher quasistationary IPS resonant states.

\section{Conclusion}

\hspace*{0.6cm} We experimentally demonstrate that for the Pb islands biased at $U\lesssim 3\,$V the change in the local thickness of the Pb film by one monolayer results in abrupt spatial variation of the differential conductance $dI/dU$ at given energy, conditioned by the QWS inside the film.  In the contrast to that, for biased voltages $U\gtrsim 3.5\,$V the energy spectrum of the modified IPS which appear above flat Pb terraces becomes thickness--independent. We estimate the local work function $W^{\,}_s=3.8\pm 0.1\,$eV for the flat Pb(111) terraces by examining the dependence of the position of $n-$th IPS resonance on its number. We show that the modification of the shape of the STM tip leads to a change of the number of the IPS resonances for the same Pb terrace, but does not affect the estimate for the local work function.

\section{Acknowlednements}

\hspace*{0.6cm} The author thanks K. Schouteden, V. S. Stolyarov, S. I. Bozhko, D. A. Ryzhov, A. V. Silhanek, and G. Karapetrov for valuable comments and stimulating discussions. The work was performed with the use of the facilities at the Common Research Center 'Physics and Technology of Micro- and Nanostructures' at Institute for Physics of Microstructures RAS. The reported study was partly funded by the Russian State Contract No. 0035-2019-0021 (sample preparation), partly funded by the Russian Fund for Basic Research, project number 19-02-00528 (STM-STS measurements and interpretation of results).

\end{document}